\documentclass[aps,twocolumn,floatfix,eqsecnum,showpacs,prb,superscriptaddress]{revtex4}
\usepackage{graphicx}
\usepackage{amsmath}
\usepackage{amssymb}
\usepackage{color}
\usepackage{bm}
\usepackage{bigstrut}
\usepackage{txfonts}
\usepackage{tabularx}

\newcolumntype{R}{>{\raggedleft\arraybackslash}X}

\begin{document}
\title{Junctions of multiple quantum wires with different Luttinger parameters}
\author{Chang-Yu Hou}
\affiliation{Department of Physics and Astronomy, University of California at Riverside, Riverside, CA 92521}
\affiliation{Department of Physics, California Institute of Technology, Pasadena, CA 91125}
\author{Armin Rahmani}
\affiliation{Theoretical Division,T-4 and CNLS, Los Alamos National Laboratory, Los Alamos, New Mexico 87545, USA}
\author{Adrian E. Feiguin}
\affiliation{Department of Physics and Astronomy, University of Wyoming, Laramie, Wyoming 82071, USA}
\author{Claudio Chamon}
\affiliation{Physics Department, Boston University, Boston, MA 02215, USA}

\date{\today}

\begin{abstract}
Within the framework of boundary conformal field theory, we evaluate the conductance of stable fixed points of junctions of two and three quantum wires with different Luttinger parameters. For two wires, the physical properties are governed by a single effective Luttinger parameter for each of the charge and spin sectors. We present numerical density-matrix-renormalization-group calculations of the conductance of a junction of two chains of interacting spinless fermions with different interaction strengths, obtained using a recently developed method [Phys. Rev. Lett. \textbf{105}, 226803 (2010)]. The numerical results show very good agreement with the analytical predictions. For three spinless wires (i.e., a Y junction) we analytically determine the full phase diagram, and compute all fixed-point conductances as a function of the three Luttinger parameters.
\end{abstract}

\maketitle

\section{Introduction}
\label{sec:introduction}

Conducting quantum wires, at low energies, generically form a Tomonaga-Luttinger liquid (TLL), characterized by a Luttinger parameter $g$, which encodes the effects of electron-electron interactions.~\cite{Tomonaga50,Luttinger63,Mattis65,Haldane81} Due to the prominent role of interactions in low dimensions, the nature of these one-dimensional electronic systems is dramatically different from their higher-dimensional counterparts described by Landau's Fermi-liquid theory.~\cite{Schulz95} The TLL state of matter in one-dimensional quantum wires has been realized in numerous experiments over the last few years.~\cite{Yacoby96,Bockrath97,Papadopoulos00,Fuhrer00,Liang01,Kim01,Terrones02,Javey03}

Transport properties of such quantum wires are of considerable interest: From a fundamental point of view, a large number of interesting phenomena have been predicted and observed. For instance, at low temperature and low bias voltage, a TLL with repulsive interactions ($g<1$) is totally disconnected in the presence of an impurity, while one with attractive interactions ($g>1$) conducts as in the absence of the  impurity.~\cite{Kane92a,Kane92b,Furusaki93,Yue94,Wong94,Frojdh95} From a practical viewpoint, junctions of TLL wires serve as important building blocks of quantum circuits,~\cite{Nayak99,Lal02} and are thus of technological significance. Junctions of three quantum wires, known as Y junctions, also have highly nontrivial transport properties.~\cite{Lal02,Chamon03} Due to their rich transport behavior, junctions of quantum wires and their networks have thus attracted much attention. ~\cite{Chen02,Pham03,Rao04,Kazymyrenko05,Oshikawa06,Bellazzini07,Hou08,Das08,Bellazzini09a,Bellazzini09b,Safi09,Aristov10,Mintchev11,Aristov11,Caudrelier12}  

Most of the previous works on the transport properties of junctions of TLL wires focus on wires with the \textit{same} Luttinger parameter. However, experimentally, there is no reason for all the TLLs emanating from a junction to be identical. Moreover, a single TLL can have inhomogeneities; for example, a contact between an interacting TLL and a Fermi-liquid lead, a key ingredient of most transport measurements, is often studied as an inhomogeneous TLL wire smoothly interpolating between interacting (TLL) and noninteracting (Fermi-liquid) regions or as a two-wire junction with the Luttinger parameter abruptly changing at the junction.~\cite{Maslov95,Safi95a,Ponomarenko95,Safi95b,Furusaki96,Chamon97,Safi99b,Lal01,Imura02,Dolcini03,Crepieux03,Enss05,Dolcini05,Janzen06,Rech08,Guigou09,Pugnetti09,Thomale11,Sedlmayr12} 
A junction of three quantum wires with different Luttinger parameters has been studied in the weak coupling regime.~\cite{Lal02,Das04,Aristov11b,Aristov12} The experimental importance of junctions of TLL wires with generally unequal Luttinger parameters motivates an in-depth study of their properties, which is the main objective of the present paper.

\begin{figure}
\includegraphics[width=6.5cm]{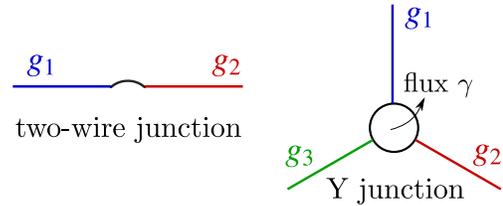}
\caption{Junctions of two and three TLLs with different Luttinger parameters.}\label{fig:fig1}
\end{figure}

Here, we focus on junctions of two and three nonchiral Luttinger liquids schematically depicted in Fig.~\ref{fig:fig1}. For two wires, it is known that the transport properties of the junction are fully controlled by one effective Luttinger parameter $g_e=2/(g_1^{-1}+g_2^{-1})$ as found in Ref.~\onlinecite{Safi95a}. In the context of fractional Hall edge states~\cite{Feiguin08}, similar results have been found for tunneling between two chiral-TLL edge states.~\cite{Chamon97}. For two nonchiral wires, one can reach the same conclusion through an almost identical argument. In this work, however, we obtain this result within the framework of boundary conformal field theory (BCFT), using the delayed evaluation of boundary conditions (DEBC) method,~\cite{Cardy89,Cardy91,Affleck95,Affleck08} which, as we will see, has the advantage that it can be readily generalized to junctions of more quantum wires. 

Such generalization to a junction of three nonidentical quantum wires is a key result of this paper. We find the stability regions of the previously identified (in Ref.~\onlinecite{Oshikawa06} for three equal Luttinger parameters) fixed points of such a Y junction in the $(g_1,g_2,g_3)$ space, and compute their corresponding conductances as a function of these three Luttinger parameters. Moreover, we obtain new asymmetric fixed points, which are only stable for nonidentical TLLs, thereby providing a more complete classification of the conformally invariant BCs for three TLLs. Such asymmetric fixed points have been identified using perturbative renormalization-group analysis in the weak coupling regime.~\cite{Lal02,Das04,Aristov11b}

Another important result of this paper is a direct numerical verification, through DMRG computations,~\cite{White92} of the analytical predictions for the conductance of a junction of two nonidentical wires. Using a recently developed method,~\cite{Rahmani10,Rahmani12} which allows us to extract the conductance from a ground-state calculation in a finite system, we compute, in a microscopic lattice model, the conductance of a junction of two chains of interacting spinless electrons with different interaction strengths. Our numerical results show excellent agreement with the DEBC predictions.

The outline of this paper is as follows. In Sec.~\ref{sec:model}, we set up the notation and present the model in the bosonization framework. In Sec.~\ref{sec:two-wire}, we present the DEBC analysis of a junction of two spinless wires and show that the scaling behavior of such junctions is governed by a single effective Luttinger parameter $g_e=2/(g_1^{-1}+g_2^{-1})$. For wires with spin-$1/2$ electrons, we obtains two effective Luttinger parameters $g_{e}^{c,s}=2/(1/g^{c,s}_1+1/g_2^{c,s})$, corresponding to charge and spin sectors. Section~\ref{sec:DMRG} contains the numerical DMRG calculations of the conductance of a junction of two  spinless chains with different interaction strengths. In Sec.~\ref{sec:3-wire}, we summarize our results on the Y junction with the detailed DEBC analysis presented in Appendix~\ref{app:3-wires}.
In Sec.~\ref{sec:G-renormalization}, we present the analysis of the conductance renormalization when the wires are contacted to Fermi-liquid leads. Finally, we conclude in Sec.~\ref{sec:conclusion}.

\section{General setup}
\label{sec:model}

In this section, we present the model in the bosonization framework and set up the notation. In the low-energy limit, the wires are described by TLLs with the Euclidean action:~\cite{Kane92a,Kane92b}
\begin{equation}
\label{eq:Bulk-action}
\begin{split}
S=& \sum_{i}  \frac{v_i g_i}{4 \pi} \int d\tau \int_0^\infty dx \left[ (\partial_x \varphi_i )^2 + \frac{(\partial_\tau \varphi_i )^2}{v_i^2} \right]
\\
=& \sum_{i}  \frac{ v_i }{4 \pi g_i} \int d\tau \int_0^\infty dx \left[(\partial_{x} \theta_i)^2 + \frac{(\partial_{\tau} \theta_i)^2}{v_i^2} \right],
\end{split}
\end{equation}
where $g^{\ }_{i}$ and $v_i$ are, respectively, the Luttinger parameter and the plasmon velocity of wire $i$. Different wires can have different electron-electron interactions and consequently different Luttinger parameters $g_i$. The boson fields, $\varphi_i$ and $\theta_i$, have the following equal-time commutation relation,
\begin{equation}
[\varphi_i(x), \theta_{j} (x')]= i \pi \delta_{ij} \mathrm{sgn}(x'-x),
\end{equation}
so the conjugate momenta of $\theta_i$ fields are given by $\Pi_{\theta_i} = (\partial_x \varphi_i)/2 \pi$. Let us also define left- and right-moving boson fields as
\begin{equation}
\label{eq:LR-boson-varphi-theta}
\varphi_{i}= \phi^{L}_i + \phi^{R}_i ; \qquad \theta_{i} = \phi^{L}_i - \phi^{R}_i.
\end{equation}

The fermions $ \psi_{i} (x) = e^{i k_F x} \psi_i^{R}(x) + e^{- i k_F x} \psi_i^{L}(x)$, with $\psi_i^{L}(x)$ and $\psi_i^{R}(x)$ the linearized left- and right-moving fermionic fields, can be written in terms of the above bosons through:
\begin{equation}
\label{eq:fermion-boson}
\psi_i^{L,R}= \frac{\eta_i}{ \sqrt{2\pi} } e^{i \sqrt{2} \phi_{i}^{L,R}} = \frac{\eta_i}{\sqrt{2\pi}}e^{i (\varphi_i \pm \theta_i)/\sqrt{2} },
\end{equation}
where $\eta_{i}$ are anticommuting Klein factors, which ensure the correct fermionic statistics. The Klein factors play no role in our analysis and are hence neglected throughout the paper. It is convenient to define complex variables $z=\tau + i x$ and $\bar z=\tau - i x$ such that the left- and right-moving bosons, corresponding to  current flowing toward and away from the junction, are, respectively, functions of $z$ and $\bar{z}$ only. The chiral current operators can then be written as
\begin{equation}
\label{eq:L-R-currents}
J_{i}^{R}= \frac{i}{\sqrt{2}\pi} \partial_{\bar{z}} \theta_i; \qquad J_{i}^{L}= \frac{-i}{\sqrt{2}\pi} \partial_{z} \theta_i,
\end{equation}
where $\partial_z=(\partial_\tau - i \partial_x)/2$ and $\partial_{\bar{z}}=(\partial_\tau + i \partial_x)/2$. The total current is proportional to the difference between the right and the left currents: $J_i=v_i(J_i^{R}-J_i^{L})$.

To analyze the junctions of quantum wires with unequal Luttinger parameters, it is convenient to introduce rescaled bosonic fields
\begin{equation}
\label{eq:def-rescaled-field}
\tilde{\theta}_i \equiv  \theta_i /\sqrt{g_i} ; \qquad \tilde{\varphi}_i \equiv  \sqrt{g_i} \varphi_i ,
\end{equation}
which effectively  have a noninteracting action [i.e., $g=1$ in Eq.~(\ref{eq:Bulk-action})]. Note that these rescaled fields satisfy the original commutation relations. Similarly, we define the following rescaled left- and right-moving bosonic fields:
\begin{equation}
\label{eq:LR-boson-varphi-theta-rescaled}
\tilde{\varphi}_{i}= \tilde{\phi}^{L}_i + \tilde{\phi}^{R}_i ; \qquad \tilde{\theta}_{i} = \tilde{\phi}^{L}_i - \tilde{\phi}^{R}_i,
\end{equation} in terms of which the left- and right-moving fermions become
\begin{equation}
\label{eq:fermion-boson-rescaled}
\psi_i^{L,R}=\exp\left[\frac{i}{\sqrt{2}} ( \frac{\tilde{\varphi}_{i} }{\sqrt{g_i}}
\pm \sqrt{g_i} \tilde{\theta}_{i})\right]. 
\end{equation}
In the absence of a junction (boundary), the correlation functions of the rescaled fields are given by: 
\begin{equation}
\label{eq:correlation-function-abs-boundary}
\begin{split}
\langle \tilde{\varphi}_i(z,\bar{z}) \tilde{\varphi}_j(w,\bar{w}) \rangle=& - \frac{\delta_{ij}}{2} \ln \left[ (z-w)(\bar{z}-\bar{w})\right] ,
\\
\langle \tilde{\theta}_i(z,\bar{z}) \tilde{\theta}_j(w,\bar{w}) \rangle = & - \frac{\delta_{ij}}{2}  \ln\left[ (z-w)(\bar{z}-\bar{w}) \right] ,
\\
\langle \tilde{\varphi}_i(z,\bar{z}) \tilde{\theta}_j(w,\bar{w}) \rangle = & 0.
\end{split}
\end{equation}
Imaginary time ordering is implied for all correlation functions here and throughout the paper.
Different bosonic fields above are of course uncorrelated. However, as we will show below, the presence of a junction mixes these fields and effectively reduces the independent bosonic degrees of freedom by half.

At the end of wires, $x=0$, fermions can hop between different wires. this process is described by a single particle hopping Hamiltonian
\begin{equation}
\label{eq:Hamiltonian-hopping}
H_B = -\sum_{i,j} \left[t_{ij} e^{i\alpha_{ij}} \psi^\dag_{i}(0) \psi_{j}(0)+  H.c. \right], 
\end{equation}
where $t_{ij}$ and $\alpha_{ij}$ are the strength and the phase of the hopping amplitude between wires $i$ and $j$. Without loss of generality~\cite{Chamon03,Oshikawa06} (at least for junctions of two or three wires), we only consider the symmetric case $t_{ij}=t$ in this paper. The  phases encode the distribution of magnetic fluxes at the junction, which for a junction of three or more wires can play a crucial role in the properties of some RG fixed points.~\cite{Chamon03,Oshikawa06}

\section{Junction of two wires: DEBC analysis}
\label{sec:two-wire}

In this section, we analyze the stability of RG fixed points, and compute their corresponding conductances for a junction of two TLL quantum wires with unequal Luttinger parameters. By using the DEBC method of Ref.~\onlinecite{Oshikawa06}, we show that the properties of such junctions only depend on an effective Luttinger parameter
\begin{equation}
\label{eq:geff}
g_{e}^{-1}=(g_{1}^{-1}+g_{2}^{-1})/2.
\end{equation}
Briefly, the junction of two wires is totally decoupled for $g_{e}<1$, and has a conductance $g_{e} (e^2/h)$ for $g_{e}>1$. This result is consistent with what was found in Ref.~\onlinecite{Chamon97}, where the tunneling between fractional quantum Hall edge states with different filling fractions was discussed. The DEBC method used in this paper, however, has the advantage that it can be straightforwardly generalized to junctions of more than two wires. (see Sec.\ref{sec:3-wire} and Appendix~\ref{app:3-wires}) 

\subsection{DEBC method}

For one-dimensional quantum impurity problems, one often invokes the conformal symmetry of the bulk system (Luttinger liquids in our case) and assumes that the effect of an impurity (junction), at low energies, is imposing a conformally invariant boundary condition (BC), which describes the renormalization-group (RG) fixed point. This methodology of relating the BC and RG fixed points of the system is called boundary conformal field theory (BCFT), and has proved greatly successful in the study of quantum impurity problems.~\cite{Affleck95,Affleck08}

A useful technique within the framework of BCFT is the DEBC method, which hugely simplifies the evaluation of the scaling dimensions, $\Delta_{\mathcal{O}_B}$, of boundary operators, $\mathcal{O}_B$, with a given BC.~\cite{Oshikawa06} The scaling dimension, in turn, determines the leading scaling behavior of a given operator under the RG flow, and thus govern the stability of the RG fixed points: In general, an RG fixed point (boundary condition) is stable if all boundary operators are either equivalent to identity or irrelevant $\Delta_{\mathcal{O}_B}>1$. Moreover, the conductance associated with the given fixed point can be readily computed from the BC. For a complete description of the DEBC method, we refer the reader to Refs.~\onlinecite{Oshikawa06,Hou08}. Here, we simply apply this method to a junction of two quantum wires with unequal Luttinger parameters, commenting only on some key ingredients. 

\subsection{Junctions of two quantum wires}

The first step of the DEBC method is to write an ansatz for the conformally invariant BCs describing the RG fixed points. The next step is to list all the boundary operators, which can possibly become relevant and make the fixed point unstable, and compute their scaling dimensions with such ansatz for every point in the parameter space, $g_1$ and $g_2$. If for a given ansatz, none of these boundary operators have a scaling dimension smaller than one (in some region of the parameter space known as the stability region), we have found a stable RG fixed point. In case of a junction, a natural ansatz can be expressed in terms of a rotation matrix $\mathcal{R}$ that relates outgoing to incoming bosonic fields:
\begin{equation}
\label{eq:BC-R-matrix}
\tilde{ \boldsymbol{\phi} }^R =\mathcal{R} \tilde{\boldsymbol{\phi} }^L,
\end{equation}
where $\tilde{ \boldsymbol{\phi} }^{L,R} \equiv (\tilde{\phi}^{L,R}_1,\dots, \tilde{\phi}^{L,R}_i)^{T}$ are $i$-component vector fields.

The most important boundary operators, in case of a junction of two wires, correspond to the following processes: tunneling of chiral fermions between the two wires, and backscattering within the individual wires. It is useful to introduce a compact notation for boundary operators describing the single-particle tunneling processes,  
\begin{equation}
\label{eq:def-Tjiba}
T_{ji}^{ba}\equiv {\psi_{j}^{b}}^\dag \psi^{a}_{i}|_{x=0},
\end{equation}
where $a,b=R,L$. We can then list these six fundamental boundary operators in terms of the rescaled boson fields as follows:
\begin{equation}
\label{eq:fund-BO-2wires}
\begin{split}
T_{21 (12)}^{RL}\sim & e^{\pm \frac{i}{\sqrt{2}}(\frac{\tilde{\varphi}_1}{\sqrt{g_1}}-\frac{\tilde{\varphi}_2}{\sqrt{g_2}})}
e^{\frac{i}{\sqrt{2}}(\sqrt{g_1} \tilde{\theta}_1 + \sqrt{g_2} \tilde{\theta}_2)},
\\
T_{11(22)}^{RL}\sim & 
e^{i \sqrt{2} \sqrt{g_1} \tilde{\theta}_{1(2)}},
\\
T_{21}^{LL(RR)}\sim & e^{\frac{i}{\sqrt{2}}(\frac{\tilde{\varphi}_1}{\sqrt{g_1}}-\frac{\tilde{\varphi}_2}{\sqrt{g_2}})}e^{\pm \frac{i}{\sqrt{2}}(\sqrt{g_1}\tilde{\theta}_1-\sqrt{g_2}\tilde{\theta}_2)}.
\end{split}
\end{equation}
The boundary operators corresponding to multiparticle processes are not forbidden and can be generated as higher-order perturbation processes even they are not presence in the bare Hamiltonian. In general, they can be constructed from these fundamental boundary operators and have larger scaling dimensions and are less relevant than the single-particle processes. 

All the above boundary operators have the generic form $\mathcal{O}_B \sim e^{i \boldsymbol{a} \cdot \tilde{\boldsymbol{\varphi} } + i \boldsymbol{b} \cdot \tilde{\boldsymbol{\theta} }}$, where $\boldsymbol{a}, \boldsymbol{b}$ are vectors that contain the prefactors of the $\tilde{\varphi}_i$ and $\tilde{\theta}_i$ fields. By eliminating the redundant degrees of freedom with Eq.~\eqref{eq:BC-R-matrix}, and using Eq.~\eqref{eq:correlation-function-abs-boundary}, the scaling dimension of the generic $\mathcal{O}_B$ above can then be written in terms of $\mathcal{R}$ as
\begin{equation}
\label{eq:scaling-dimension-Rab}
  \Delta^{\mathcal{R}}_{\mathcal{O}_B} = \frac{1}{4} |\mathcal{R}^{T} (\boldsymbol{a}-\boldsymbol{b}) + (\boldsymbol{a}+\boldsymbol{b})|^2,
\end{equation}
where the superscript $T$ represents matrix transpose.

To find all the $\mathcal{R}$ matrices that correspond to stable fixed points, it is convenient to express $\tilde{\varphi}_{i}$ fields in terms of the following $\pm$ fields:
\begin{equation}
 \varphi_{+} = \frac{ \sqrt{g_1} \tilde{\varphi}_1 + \sqrt{g_2} \tilde{\varphi}_2 }{\sqrt{g_1+g_2}}
; \;
 \varphi_{-} = \frac{\sqrt{g_2} \tilde{\varphi}_1 - \sqrt{g_1} \tilde{\varphi}_2 }{\sqrt{g_1+g_2}}.
\end{equation}
Corresponding ${\theta}_{\pm}$ are defined in a similar manner.
The six fundamental boundary operators can then be written as
\begin{equation}
\label{eq:fund-BO-2wires-pm-fields}
\begin{split}
T_{21 (12)}^{RL} \sim &  e^{\pm i \frac{\varphi_{-}}{\sqrt{g_e}} },
\\
T_{11(22)}^{RL} \sim &  e^{ \pm i \sqrt{g_e} \theta_{-}  },
\\
T_{21}^{LL(RR)} \sim & e^{i \frac{\varphi_{-}}{\sqrt{g_e}} } 
e^{ \pm i \sqrt{g_e} \theta_{-} } .
\end{split}
\end{equation}
Here, we have dropped all $e^{ i \theta_{+} }$ terms as they are effectively an identity at the boundary: Charge conservation requires $\sum_i J_i^R-J_i^L=0$, which, using Eqs.~(\ref{eq:L-R-currents}) and  (\ref{eq:def-rescaled-field}), gives $\partial_\tau \theta_{+}=0$ (i.e., Dirichlet BC  on $\theta_{+}$), and makes $e^{ i \theta_{+} }$ an effective identity.~\cite{Oshikawa06}
The simplified boundary operators in Eq.~\eqref{eq:fund-BO-2wires-pm-fields} then only depend the effective Luttinger liquid parameter $g_e$.
 
In terms of the left- and right-moving $\pm$ fields defined in a similar manner to Eq.~\eqref{eq:LR-boson-varphi-theta-rescaled}, the Dirichlet BC on $\theta_{+}$ gives
\begin{equation}
\label{eq:BC-+field}
\phi^{R}_{+} = \phi^{L}_{+} \big|_{x=0}.
\end{equation} 
Now we only need to specify the BC relating the $\phi^{L,R}_{-}$ fields as:
\begin{equation}
\label{eq:BC--field}
 \phi^{R}_{-} = \mathcal{R}_{-} \phi^{L}_{-} \big|_{x=0}.
\end{equation}
Because there is a single pair of $RL$ fields, only the Neumann-BC (N-BC) and the Dirichlet-BC (D-BC), $\mathcal{R}_{-}^{N,D}= \pm 1$, are allowed. By using Eq.~\eqref{eq:scaling-dimension-Rab} on boundary operators listed in Eq.~\eqref{eq:fund-BO-2wires-pm-fields}, we obtain the scaling dimension of each operator with the N-BC and D-BC:
\begin{center}
\begin{tabular}{c|c|c}
 $\mathcal{O}_B$ & $\Delta^{N}_{\mathcal{O}_B}$ (N-BC) & $\Delta^{D}_{\mathcal{O}_B}$ (D-BC) \\ 
\hline $T_{21}^{RL}$, $T_{12}^{RL}$ & $1/g_e$ & $0$ \\ 
\hline $T_{11}^{RL}$, $T_{22}^{RL}$ &  $0$ & $g_e$ \\ 
\hline $T_{21}^{LL}$, $T_{21}^{RR}$ & $1/g_e$ & $g_e$ \\  
\end{tabular}  
\end{center}
Here, the scaling dimension $0$ indicates that $\mathcal{O}_B$ is equivalent to identity operator $\openone$ for the given boundary condition.

From the table above, we conclude that the N-BC is stable when $g_e < 1$ and D-BC is stable when $g_e > 1$. As shown below, the N-BC corresponds to a fixed point where two wires are disconnected and the D-BC corresponds to a fixed point where two wires are maximally connected with conductance $G_{D}= g_e (e^{2}/h)$. Let us recall the well-known results of the Kane and Fisher problem, namely, a single impurity in a spinless TLL wire, which is equivalent to a junction of two wires with the same Luttinger parameter, $g$.~\cite{Kane92a} There, two RG fixed points are identified, totally disconnected fixed point for $g<1$ and maximally connected fixed point for $g>1$ with $G_{D}= g (e^{2}/h)$. As $g_e\to g$ when $g_1=g_2=g$, the fixed points, which we identified, generalize Kane and Fisher's results to the case of two quantum wires with different Luttinger parameters.

A similar DEBC analysis can also be applied to junction of two quantum wires with spin-$1/2$ electrons.~\cite{Hou08} In that case, we obtain two effective Luttinger parameters $g_{e}^{c,s}=2/(1/g^{c,s}_1+1/g_2^{c,s})$, corresponding to charge and spin sectors, which govern the scaling behavior of boundary operators near the fixed points.~\cite{Safi99a} As a consequence, the stable fixed points of such system are connected to those of equal Luttinger parameters but with $g^{c,s}\to g_e^{c,s}$.~\cite{Kane92b,Furusaki93,Wong94}

\subsection{Conductance for each fixed point}
\label{sec:conductance-2wire}

Here, we will compute the conductance associated with N-BC and D-BC. The conductance tensor is defined through the following current-voltage relation:  
\begin{equation}
\label{eq:G-def}
I_i= \sum_{j} G_{ij} V_{j},
\end{equation}
where current is defined as positive when flowing toward junction.
In the linear-response regime, the conductance above can be evaluated via the Kubo formula:~\cite{Oshikawa06}
\begin{multline}
\label{eq:Kubo-formula}
G_{ij}= \lim_{\omega\to 0_+} 
\\
-\frac{e^{2}}{\hbar} \frac{1}{\omega L} \int_{-\infty}^{\infty}d\tau e^{i \omega \tau} \int_{0}^{L} dx \langle J_i(y,\tau)J_j(x,0) \rangle.  
\end{multline}
The current correlation function can be rewritten as a sum of chiral-current correlation functions as
\begin{multline}
\langle J_i(y,\tau)J_j(x,0) \rangle =
\\
\langle J^{R}_i(y,\tau)J^{R}_j(x,0) \rangle 
+\langle J^{L}_i(y,\tau)J^{L}_j(x,0) \rangle
\\
-\langle J^{R}_i(y,\tau)J^{L}_j(x,0) \rangle
-\langle J^{L}_i(y,\tau)J^{R}_j(x,0) \rangle ,
\end{multline}
where the left and right currents are defined in Eq.~\eqref{eq:L-R-currents}. By using Eq.~\eqref{eq:def-rescaled-field}, we can write the chiral currents in terms of rescaled boson fields as follows:
\begin{equation}
\label{eq:rescaled-LR-current}
\begin{split}
J_{i}^{R} =& + i \frac{\sqrt{g_i}}{\sqrt{2}\pi} \partial_{\bar{z}} \tilde{\theta}_i = -i \frac{\sqrt{g_i}}{\sqrt{2}\pi} \partial_{\bar{z}} \tilde{\phi}^{R}_i \equiv \sqrt{g_i} \tilde{J}_i^{R},
\\
J_{i}^{L} =& -i\frac{\sqrt{g_i}}{\sqrt{2}\pi} \partial_{z} \tilde{\theta}_i= -i\frac{\sqrt{g_i}}{\sqrt{2}\pi} \partial_{z} \tilde{\phi}^{L}_i \equiv \sqrt{g_i} \tilde{J}_i^{L},
\end{split}
\end{equation}
where we have defined left and right currents associated with the rescaled fields.

To evaluate the above correlation functions given a BC, it is convenient to first express the boundary conditions $\mathcal{R}$ directly in the rescale boson field, $\tilde{\phi}_{i}^{L,R}$, basis. From Eq.~\eqref{eq:BC-+field} and Eq.~\eqref{eq:BC--field}, we can derive the rotation matrix:
\begin{equation}
\label{eq:R-ND-BC}
\mathcal{R}^{N}= \openone ;
\qquad
\mathcal{R}^{D} =
\left(
\begin{array}{cc}
\frac{g_1-g_2}{g_1+g_2} & \frac{2 \sqrt{g_1g_2}}{g_1+g_2} \\ 
\frac{2 \sqrt{g_1g_2}}{g_1+g_2} & \frac{g_2-g_1}{g_1+g_2} 
\end{array}
\right),
\end{equation}
for the N-BC and D-BC, respectively. From Eq.~\eqref{eq:rescaled-LR-current}, the left and right currents are constrained by the $\tilde{J}^{R}_i(0,\tau) = \mathcal{R}_{ij} \tilde{J}^{L}_j(0, \tau)$ BC at the origin, which, upon unfolding the current, implies
\begin{equation}
\label{eq:unfold-currents-relation}
 \tilde{J}^{R}_i(x,\tau) = \mathcal{R}_{ij} \tilde{J}^{L}_j(-x, \tau),
\end{equation}
for $x>0$. Therefore, all the right-moving currents can be interpreted as left-moving on the $x<0$ domain.  

Now, the chiral current correlation functions can be evaluated:
\begin{equation}
\label{eq:LR-correlation-function}
\begin{split}
\langle J^{R}_i(\bar{z}_i)J^{R}_j(\bar{z}_j) \rangle  =& \frac{\delta_{ij}}{ 4\pi^{2} } \frac{ g_i }{(\bar{z}_i -\bar{z}_j)^2} ,
\\
\langle J^{L}_i(z_i)J^{L}_j(z_j) \rangle  =& \frac{\delta_{ij}}{ 4\pi^{2} } \frac{ g_i }{(z_i -z_j)^2} ,
\\
\langle J^{R}_i(\bar{z}_i)J^{L}_j(z_j) \rangle  =& \frac{\mathcal{R}_{ij}}{ 4\pi^{2} } \frac{ \sqrt{g_i g_j} }{(\bar{z}_i -z_j)^2} ,
\\
\langle J^{L}_i(z_i)J^{R}_j(\bar{z}_j) \rangle  =& \frac{\mathcal{R}_{ji}}{ 4\pi^{2} } \frac{ \sqrt{g_i g_j} }{(z_i -\bar{z}_j)^2} .
\end{split}
\end{equation}
By inserting these correlation functions into the Kubo formula Eq.~\eqref{eq:Kubo-formula}, and after some algebra, we obtain a concise relation between conductances and boundary conditions~\cite{Oshikawa06}
\begin{equation}
\label{eq:Gij-Rij}
 G_{ij}= \frac{e^2}{h} \sqrt{g_i g_j} (\delta_{ij} - \mathcal{R}_{ij} ).
\end{equation}
With the N-BC and D-BC represented by the rotation matrices of Eq.~\eqref{eq:R-ND-BC}, we immediately conclude that
\begin{equation}
G^{N}= 0;\qquad G^{D}= g_e \frac{e^2}{h}
\left(
\begin{array}{cc}
1 & -1 \\ 
-1 & 1
\end{array}
\right). 
\end{equation}
As expected, the N-BC corresponds to a fixed point with decoupled wires and D-BC corresponds to a fixed point with conductance $g_e (e^2/h)$.

Here, we shall emphasize that the correlation functions listed in Eq.~\eqref{eq:LR-correlation-function} include only the universal part for a given boundary condition. There are also nonuniversal contributions to the correlation functions, which, in general, decay faster and become irrelevant at long distance. However, when the universal part vanishes, which is the case for correlations between different wires with the N-BC, the higher-order contributions dominate and could lead to nonlinear conductance. In the next section, we will perform the DMRG calculations and confirm that Eq.~\eqref{eq:LR-correlation-function} indeed represents the universal part of chiral-current correlation functions.

\section{Junction of two wires: DMRG calculations}
\label{sec:DMRG}

\begin{figure}
\includegraphics[width=7cm]{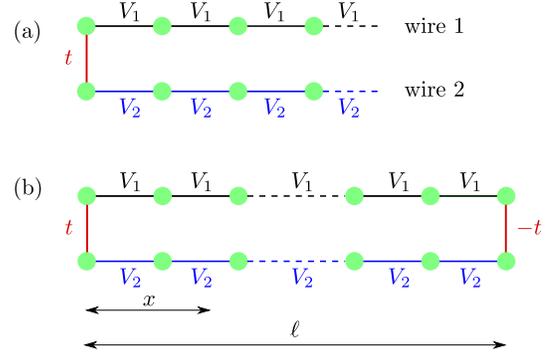}
\caption{a) The junction of two semi-infinite Luttinger-liquid wires with interaction strength $V_1$ and $V_2$. Hopping amplitude is set to unity in the bulk of the the two wires, and is equal to $t$ at the junction. b) The corresponding finite system used for DMRG calculations according to the method of Refs.~\cite{Rahmani10, Rahmani12}. }\label{fig:fig2}
\end{figure}

In this section, we perform numerical computations of the conductance of a junction of two Luttinger liquids with different Luttinger parameters $g_1$ and $g_2$.
A microscopic lattice model with the Luttinger-liquid physics, which is suitable for numerical calculations, is the one-dimensional tight-binding model of interacting spinless electrons,
\begin{equation}
\label{eq:Hubbard}
H_i=\sum_m c^\dagger_{i, m} c_{i ,m+1}+{\rm H.c.}+V_i(n_{i, m}-{1 \over 2})(n_{i, m+1}-{1 \over 2}),
\end{equation}
for wire $i$, where the hopping amplitude is set to unity. A junction of two wires can be described by
\begin{equation}\label{eq:lattice_two_wire}
H=H_1+H_2-t  c^\dagger_{1, 0}c_{2 ,0}-tc^\dagger_{2, 0}c_{1 ,0}.
\end{equation}
The parameters $g_i$ and $v_i$ of the Luttinger-liquid Hamiltonian~\eqref{eq:Bulk-action} are related to the interaction strength $V$ through the Bethe ansatz (see, e.g., Refs.~\onlinecite{Rahmani10,Giamarchi}).
We use the method of Refs.~\onlinecite{Rahmani10, Rahmani12} to compute the conductance. This method allows us to extract the conductance from a ground-state static calculation in a finite system as explained below. The semi-infinite junction of Hamiltonian~(\ref{eq:lattice_two_wire}) is depicted in Fig.~\ref{fig:fig2}(a), while the corresponding finite system used in the numerics is shown in Fig.~\ref{fig:fig2}(b). 

We extract the conductance $G_{ij}$ from the following asymptotic (large $x$) relationship in the finite system shown in Fig.~\ref{fig:fig2}(b):
\begin{equation}\label {eq:key_relation}
\langle J_R^i(x)J_L^j(x) \rangle\simeq  \frac{h}{e^2}G_{ij}\left[4\: \ell\:
\sin\left( \frac{\pi}{\ell}x\right) 
\right]^{-2},
\end{equation}
where $\langle J_R^i(x)J_L^j(x)\rangle$ is the ground-state correlation function of chiral currents in wires $i$ and $j$ in Fig.~\ref{fig:fig2}b. In terms of total charge $N$ and current $J$ operators, which can be modeled on the lattice, we generically have~\cite{Rahmani12}
\begin{equation}\label{eq:measure}
\langle J_R^i (x)J_L^j(x)\rangle=-\dfrac{1}{2v_i v_j}\langle J^i(x)J^j (x)\rangle-\dfrac{1}{2v_j}\langle N^i(x)J^j(x)\rangle.
\end{equation}
Note that in the time-reversal symmetric case considered here, the second term in the expression above vanishes, and we only need to compute a static current-current correlation function $\langle J^i(x)J^j (x)\rangle$. In terms of the lattice creation and annihilation operators appearing in Hamiltonian~(\ref{eq:Hubbard}), we have
\[
J^i (m+\dfrac{1}{2})=i({c}^\dagger_{i,m+1}c_{i,m}-{c}^\dagger_{i,m}c_{i,m+1}).
\]
All we need to do now is to numerically compute $\langle J^1(x)J^2(x)\rangle$ for the above current operator, and divide it by $2v_1v_2$ to obtain $\langle J_R^1(x)J_L^2 (x)\rangle$. The numerical calculations are done for system of $180$ sites in each of the two wires. The truncated number of states in our DMRG computations is $1100$.

\subsection {Repulsive effective interaction $g_e <1$}

For $g_e <1$, we have $G_{12}=0$, which implies that the leading term [Eq.~(\ref{eq:key_relation})] in the $\langle J_R^1(x)J_L^2(x)\rangle$ correlation function vanishes. If, as a function of ${\ell \over \pi} \sin\left({\pi \over \ell} x\right)$, the computed $\langle J_R^1(x)J_L^2(x)\rangle$ decays faster than a power law with exponent $-2$, we have a signature of a vanishing $G_{12}$. Our numerical results indeed confirm this: for any combination of $g_1$ and $g_2$ with $g_e<1$, we find that $\langle J_R^1(x)J_L^2(x)\rangle$ decays as ${\ell \over \pi} \sin\left({\pi \over \ell} x\right)^{-\alpha(g_e)}$ with $\alpha(g_e)>2$. The exponent $\alpha$ only depends on $g_e$, and is independent of individual $g_i$ and the hopping amplitude $t$. The prefactor of the correlation function depends on the hopping amplitude $t$ since universality is a property of the leading term, and the coefficient of the subleading term observed here can depend on microscopic details such as $t$. 

Note that the correlation functions for different combinations of $g_1$ and $g_2$, which have the same $g_e$, collapse not only in the large $x$ limit but also close to the microscopic length scales. This strongly suggests that the single parameter $g_e$ determines \textit{all} the subleading corrections to the correlation function. This behavior can be understood by noting that all the boundary operators in Eq.~\eqref{eq:fund-BO-2wires-pm-fields} depend on $g_e$ (as opposed to individual $g_1$ and $g_2$) after dropping the $\tilde{\varphi}_+$ and $\tilde{\theta}_+$ fields due to current conservation. Hence, the measured correlation functions should also be determined only by the effective Luttinger parameter $g_e$ and the hopping strength $t$. Interestingly, the exponent $\alpha$ is indeed close to the scaling dimension of the leading irrelevant operator (i.e., $2/g_e$) but we are not able to make a definitive statement due to finite-size effects and limited numerical precision. It is worth mentioning that we have also considered combinations of $g_1$ and $g_2$ where one wire has attractive interactions ($g_1>1$), but this does not affect the behavior of the junction as long as $g_e <1$.

\begin{figure}
\includegraphics[width=7.5cm]{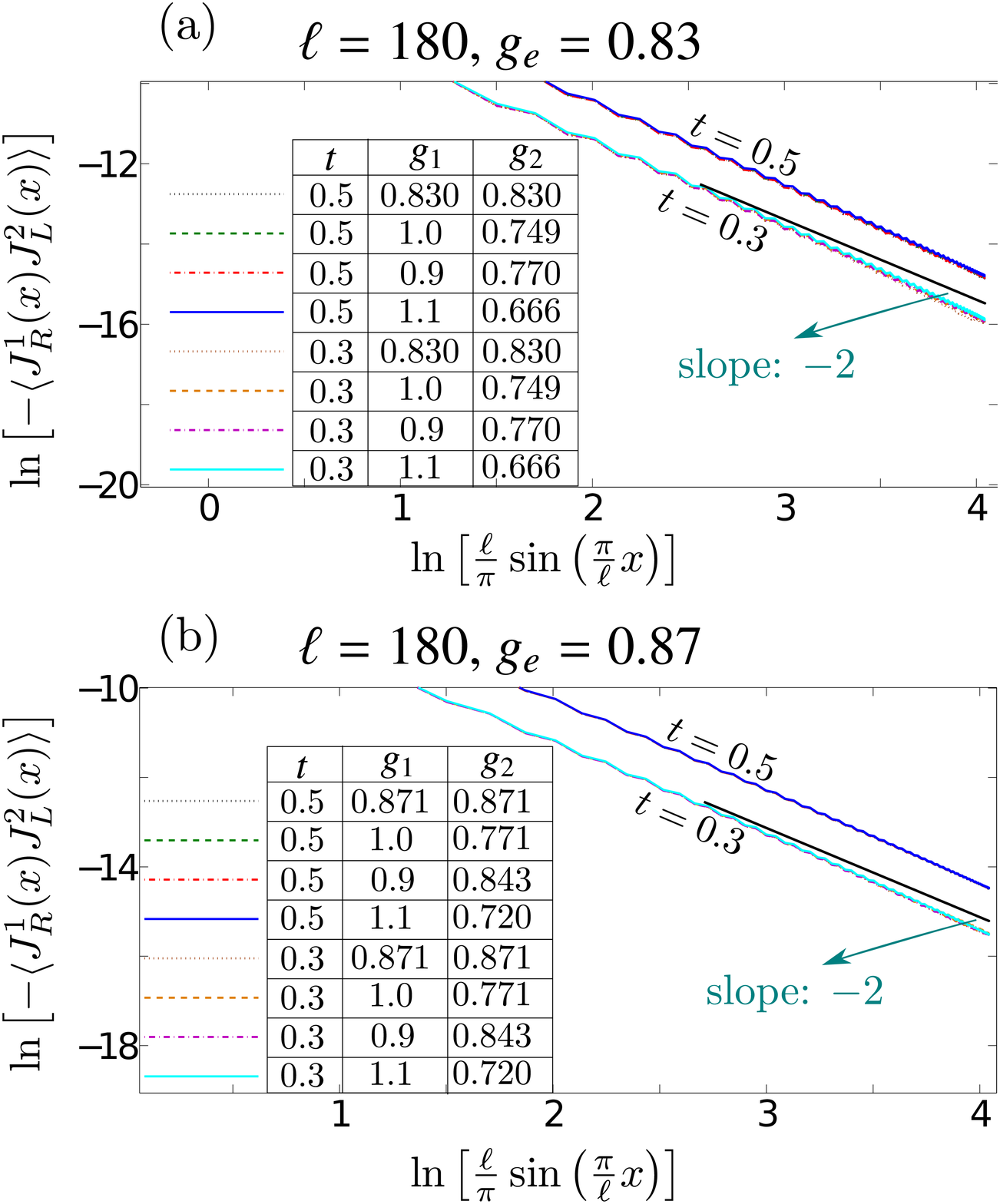}
\caption {The correlation function $\langle J_R^1(x)J_L^2 (x)\rangle$ for different combinations of $g_1$ and $g_2$ corresponding to two values of $g_e<1$. The data exhibits very good collapse, and indicates a vanishing conductance.
}\label{fig:repulsive}
\end{figure}

The results are summarized in Fig.~\ref{fig:repulsive}. We have considered two values of $g_e=0.871,0.83$ and four combinations of $g_1$ and $g_2$ for each $g_e$ as shown below.
\begin{center}
\begin{small}
\begin{tabular*}{\linewidth}{@{\extracolsep{\fill}}*{5}{c}}
\hline \hline
 $V_1$ & $V_2$ & $g_1$ & $g_2$ &$g_e$  \bigstrut 
\\ 
\hline
$0.463$ & $0.463$ & $0.871$& $0.871$& $0.871$\bigstrut
\\
$0$ & $0.9$ & $1$& $0.771$& $0.871$\bigstrut
\\
$0.347$ & $0.576$ & $0.9$& $0.843$& $0.871$\bigstrut
\\
$-0.285$ & $1.145$ & $1.1$& $0.720$& $0.871$\bigstrut
\\
\hline      
$0.632$ & $0.632$ & $0.830$& $0.830$& $0.830$\bigstrut
\\   
$0$ & $1.2$ & $1$& $0.709$& $0.830$\bigstrut
\\ 
$0.347$ & $0.904$ & $0.9$& $0.770$& $0.830$\bigstrut
\\
$-0.285$ & $1.415$ & $1.1$& $0.666$& $0.830$    \bigstrut
\\
\hline \hline
\end{tabular*}
\end{small}
\end{center}
Here, we have considered two different values of hopping amplitude, $t=0.5,0.7$.

\subsection {Attractive effective interaction $g_e >1$}

\begin{figure}
\includegraphics[width=7.5cm]{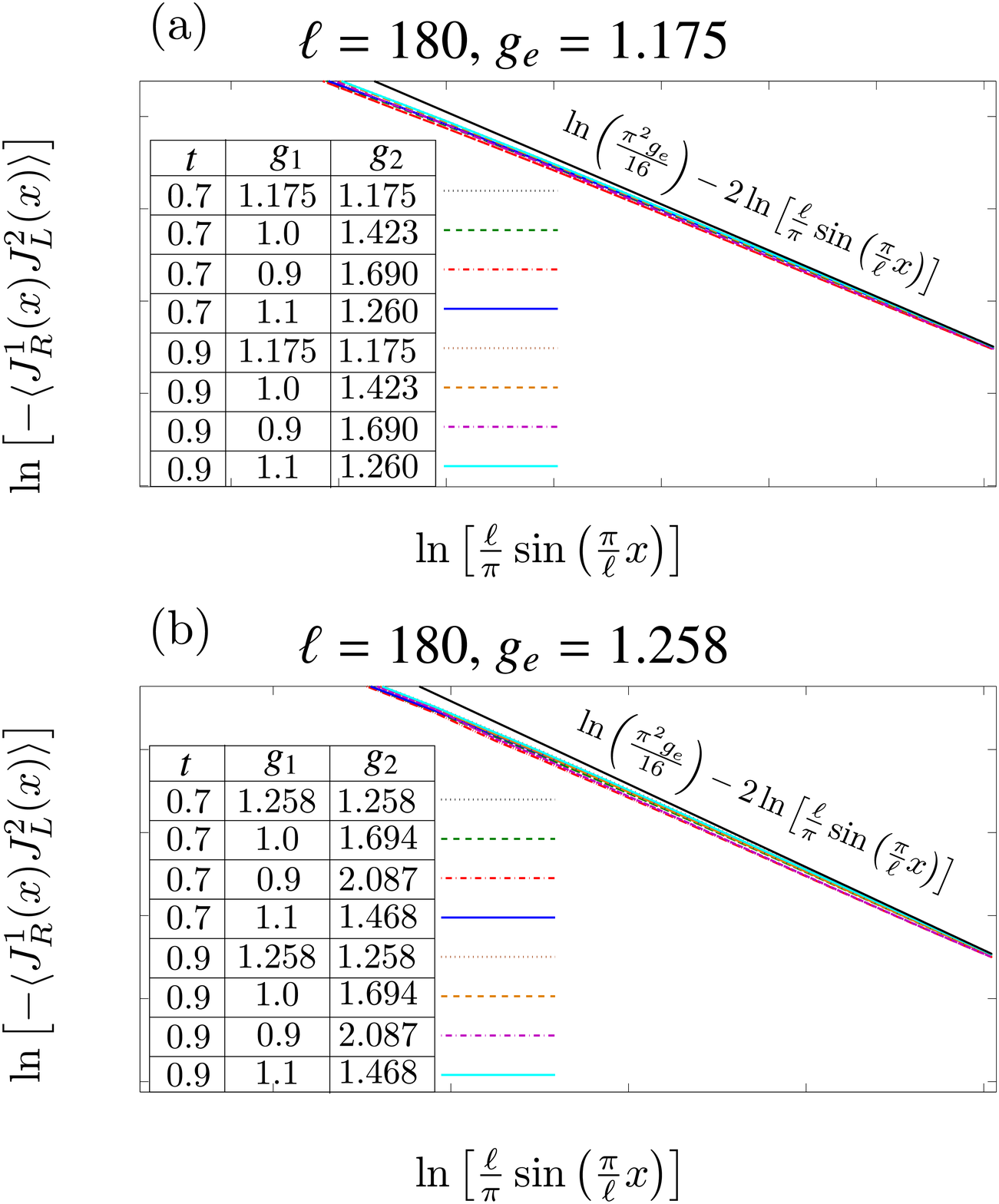}
\caption {The correlation function $\langle J_R^1(x)J_L^2 (x)\rangle$ for different combinations of $g_1$ and $g_2$ corresponding to two values of $g_e>1$. The data exhibits very good collapse, and indicates a conductance of $g_e e^2/h$ because of the asymptotic agreement with the exact theoretical prediction shown by a sold black line. }\label{fig:attractive}
\end{figure}

In this case, we expect any tunneling amplitude $t \neq 1$ at the junction to heal, and result in a universal conductance $g_e e^2/h$. In our numerics, we have used two values of tunneling amplitude, $t=0.7, 0.9$. Similarly to the repulsive case, we consider two values of $g_e=1.175,1.258$; for each $g_e$, we consider four combinations of $g_1$ and $g_2$ shown below.
\begin{center}
\begin{small}
\begin{tabular*}{\linewidth}{@{\extracolsep{\fill}}*{5}{c}}
\hline \hline
$V_1$ & $V_2$ & $g_1$ & $g_2$ &$g_e$ \bigstrut 
\\ 
\hline
$-0.4625$ & $-0.4625$ & $1.175$ & $1.175$ &$1.175$ \bigstrut
\\
$0$ & $-0.9$ & $1$ & $1.423$ &$1.175$ \bigstrut
\\
$0.347$ & $-1.196$ & $0.9$ & $1.690$ &$1.175$ \bigstrut
\\ 
$-0.285$ & $-0.637$ & $1.1$ & $1.260$ &$1.175$ \bigstrut
\\
\hline
$ -0.632$ &  $-0.632$&  $1.258$ &$1.258$ & $1.258$  \bigstrut
\\
$ 0$ &  $-1.2$&  $1$ &$1.690$ & $1.258$\bigstrut
\\
$ 0.347$ &  $-1.460$&  $0.9$ &$2.087$ & $1.258$\bigstrut
\\
$ -0.285$ &  $-0.960$&  $1.1$ &$1.468$ & $1.258$\bigstrut
\\
\hline \hline
\end{tabular*}
\end{small}
\end{center}
Again, we obtain very good agreement with the analytical predictions. The results are summarized in Fig.~\ref{fig:attractive}. With fixed $g_e$, the data points collapse for different combinations of $g_1$ and $g_2$ and different values of hopping $t$. The numerical results approach the analytical prediction for a conductance of $g_e e^2/h$ (solid black line) in the asymptotic limit.

\section{Junction of three wires}
\label{sec:3-wire}
A junction of three quantum wires with equal Luttinger parameters has three distinct types of fixed points described by a rotation matrix ansatz: decoupled fixed point, chiral-$\chi_{\pm}$ fixed points, and Dirichlet fixed points, which are respectively stable for $g<1$, $1<g<3$ and $g>3$. (There is an additional less understood time-reversal-invariant M fixed point for $1<g<3$, which we do not consider in this work.)~\cite{Chamon03,Oshikawa06,Rahmani10,Rahmani12} Thus, it is reasonable to expect that these fixed points remain stable if we slightly change the Luttinger parameters and make them unequal. Here, we first determine this region of stability around the $g_1=g_2=g_3$ line. In addition, we identify three asymmetric fixed points, only realized for unequal Luttinger parameters, in which one of the wires is decoupled from the junction and the other two wires are fully connected. These asymmetric fixed points have important consequences for the stability of N, D, and $\chi$ ones: there are regions of the parameter space near the transition points $g_i=1,3$ on the $g_1=g_2=g_3$ line where small perturbations \textit{normal} to the equal-$g$ line would drive the system into one of the asymmetric fixed points.

In Table~\ref{tab:summary-Y-junction}, we first summarize the scaling dimensions of the leading boundary operators as well as their corresponding conductances as a function of the three Luttinger parameters for each boundary condition. These scaling dimensions determine the stability of the fixed point: when all them are larger than one in certain parameter region, the given fixed point is stable. From Table~\ref{tab:summary-Y-junction}, we observe that the three scaling dimensions of the leading irrelevant operators for the asymmetric fixed points are each the inverse of those of the decoupled, Dirichlet and chiral fixed points. Hence, in any given point of the $(g_1,g_2,g_3)$ parameter space, there exists at least one stable fixed point. In other words, the decoupled, Dirichlet, chiral and asymmetric fixed points fully cover the phase diagram of a Y junction of spinless TLL wires. 

In this paper, the DMRG analysis is not applied to the junction of three quantum wires with unequal Luttinger parameters. Such analysis with equal Luttinger parameters was performed in Ref.~\onlinecite{Rahmani12}, but is  beyond the scope of the present paper. We discuss the physical properties of each of these stable fixed points in the remainder of this section. The detailed analysis, based on the DEBC method, can be found in Appendix~\ref{app:3-wires}. As the conductances of fixed points can be evaluated in the similar way as in Sec.~\ref{sec:conductance-2wire}, we simply write down the results in the following discussion.

\begin{center}
\begin{table*}
\caption{\label{tab:summary-Y-junction} The scaling dimensions of leading irrelevant boundary operators and the conductance tensor for each stable fixed point of a Y junction. The detailed analysis for obtaining these scaling dimensions is given in Appendix~\ref{app:3-wires} with the corresponding operators listed in Table~\ref{tab:Y-BO}. The asymmetric fixed point $A_i$ represents a boundary condition where the wire $i$, for $i=1,2,3$, is decoupled from the junction. Here, the scaling dimensions of all leading irrelevant operators run over the indices for all possible combinations of $i \neq j\neq k$. We have also introduced following notations: the cyclic identification $g_0 \equiv g_3$ and $g_4 \equiv g_1$; the two indices antisymmetric tensor $\epsilon_{j,j\pm 1}=\pm 1$ and $0$ otherwise; and the index $m$ satisfies $m\neq j\neq k$. The conductance tensors are given units of $e^2/h$ and are defined through $I_j= \sum_{k} G_{jk} V_k$.}
\begin{tabular*}{\textwidth}{l@{\extracolsep{\fill}}c@{\extracolsep{\fill}}c}
\hline \hline
Fixed point & Scaling dimensions $\Delta$  & Conductance $G_{jk}$ [$e^2/h$] \bigstrut
\\ 
\hline
Decoupled (N-BC) & $(g_i+g_j)/2g_i g_j$ & $0$ \bigstrut
\\ 
\hline
Dirichlet (D-BC)  
&
$g_i(g_j+g_k)/2(g_1+g_2+g_3)$  
& 
$2[ g_j \delta_{jk} -  g_j g_k /(g_1+g_2+g_3)]$ 
\bigstrut
\\ 
\hline
 Chiral ($\chi_{\pm}$-BC) & $ 2 g_i (g_j+g_k)/(g_1 g_2 g_3+g_1+g_2+g_3)$ &  $2 \frac{ g_j( g_1+g_{2}+g_{3}) \delta_{jk} + g_j g_k (\mp g_m \epsilon_{jk} -1 )}{g_1g_2g_3+g_1+g_2+g_3} $ 
 \bigstrut
\\ 
\hline
  & $(g_1+g_2+g_3+g_1g_2g_3)/2 g_i(g_{i+1}+g_{i-1})$ & 
\\
Asymmetric ($A_i$-BC) &
$2 g_{i+1} g_{i-1}/(g_{i+1}+g_{i-1})$ & $ \frac{2g_{i+1} g_{i-1}}{g_{i+1}+g_{i-1} } (-1 +\delta_{ij}+ \delta_{ik}+ 2\delta_{jk} -3\delta_{ij}\delta_{ik})$
\\
 & $ 2 (g_1+g_2+g_3)/g_i(g_{i+1}+g_{i-1}) $ & \bigstrut
\\ 
\hline \hline
\end{tabular*}
\end{table*}
\end{center}

\begin{center}
\textbf{a. Decoupled fixed point}
\end{center}

\begin{figure}
\includegraphics[width=7cm]{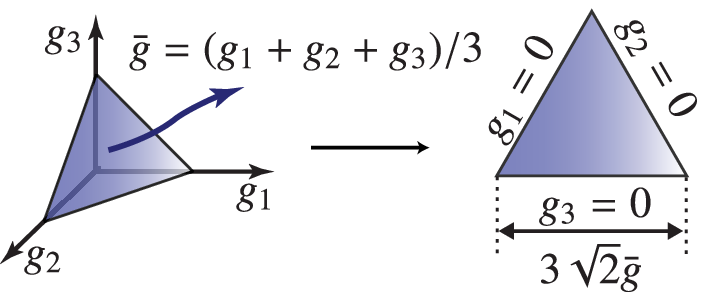}
\includegraphics[width=7cm]{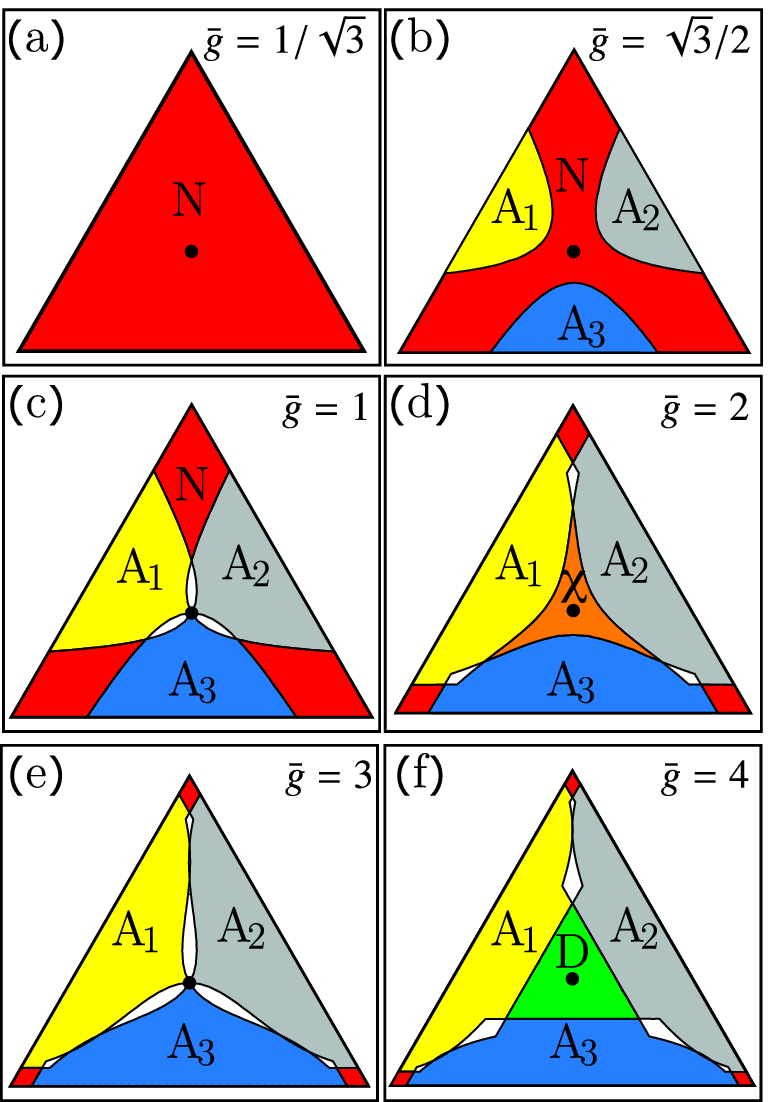}
\caption{We plot the stability regions of the different fixed points in the triangular cross sections, shown in upper panel, of fixed $\bar{g}=(g_1+g_2+g_3)/3$ in Luttinger parameters space, $g_1$, $g_2$, and $g_{3}$. In all panels, the decoupled, chiral and Dirichlet fixed points are painted, respectively, in red, orange, and green, and, the asymmetric fixed points $A_{1,2,3}$ are pained in yellow, gray, and blue, respectively. The white areas represent an overlap of two Asymmetric fixed points and the equal-$g$ points are indicated by a black dot at the center of triangle. (a) For $\bar{g}<2/3$, only the decoupled fixed point is realized. (b) The $\chi_{\pm}$ and Dirichlet fixed points are not stable when $\bar{g}<1$. (c) The point $g_{1,2,3}=1$ is an exactly marginal point surrounded by asymmetric fixed points. (d) The $\chi_{\pm}$ fixed points appear at the center of cross section for $1<\bar{g}<3$, surrounded by asymmetric fixed points. The stability region for the N-BC are pushed to the corners where two of the three Luttinger parameters become much less than one. (e) The point $g_{1,2,3}=3$ is another exactly marginal point, again surrounded by asymmetric fixed points. The chiral fixed points have extremely small stability regions (difficult to see in the figure) located between any two asymmetric fixed points. (f) The Dirichlet fixed point appears at the center of triangle for $\bar{g}>3$. Note that the figure is not a schematic, and represents the exact domain boundaries for the shown $\bar{g}$.
}
\label{fig:3-wires-fixed-point}
\end{figure}

The decoupled fixed point corresponds to the Neumann BC for all the bosonic fields $\tilde{\varphi}$. From Table~\ref{tab:summary-Y-junction}, we see that the scaling dimensions of the leading irrelevant operators are equal to $1/g_{e}^{i,j}$ for $g_e^{i,j}= 2g_i g_j/ (g_i+g_j) $, which is the same as Eq.~\eqref{eq:geff} for a pair of wires. Hence, the decoupled fixed point (N-BC) is then stable when the N fixed point is stable for all three possible pairs of wires. One can simply check that these scaling dimensions reduce to $1/g$ when the Luttinger parameters are all equal. In Fig.~\ref{fig:3-wires-fixed-point}, the stability region of the decoupled fixed point, $\Delta^{N}_{\mathcal{O}_B}>1$, is shown in red. As expected, the conductance of the decoupled fixed points is simply
\begin{equation}
G_{ij}^{N}= 0.
\end{equation}

\begin{center}
\textbf{b. Dirichlet fixed point}
\end{center}

As discussed in Appendix~\ref{app:3-wires}, one can construct three independent linear combinations of the bosonic field $\tilde{\varphi}$ such that, akin to $\tilde{\varphi}_{+}$ field for junctions of two wires, one of them, known as the center of mass field, always satisfies the Neumann BC due to charge conservation. The Dirichlet fixed point corresponds to imposing the D-BC on the other two combinations. None of single particle processes becomes identity with such boundary condition. Instead, some of two- or more-particle processes reduce to identity, which suggests that the Dirichlet fixed point is associated with a certain type of ``Andreev'' reflection that enhances the conductance.

From Table~\ref{tab:summary-Y-junction}, the scaling dimensions of all leading irrelevant operators, $
\Delta_{\mathcal{O}_B}^D= g_i(g_j+g_k)/2(g_1+g_2+g_3)$ $\forall$ $i\neq j\neq k$, reduce to $g/3$ when $g_1=g_2=g_3=g$. Hence, the D-BC becomes stable at $g>3$, consistent with Ref.~\onlinecite{Oshikawa06}. By requiring $\Delta_{\mathcal{O}_B}^{D} >1$, we obtain the stability region, shown in green in Fig.~\ref{fig:3-wires-fixed-point}, of the Dirichlet fixed point for unequal Luttinger parameters.

The conductance of the Dirichlet fixed point is given by
\begin{equation}
G_{jk}^D= 2
\frac{e^2}{h} \left[g_j \delta_{jk} - \frac{ g_j g_k  }{g_1+g_2+g_3} \right] ,
\end{equation}
where we have made a cyclic identification $g_0 \equiv g_3$ and $g_4 \equiv g_1$. When all the Luttinger parameters are equal, we have
$G_{jk}^D = g (e^2/h) (2 \delta_{jk}-2/3)$, which reproduces the result in Ref.~\onlinecite{Oshikawa06}.

\begin{center}
\textbf{c. Chiral-$\chi_{\pm}$ fixed points}
\end{center}

The chiral-$\chi_{\pm}$ fixed points have a particular transport feature: the realization of $\chi_{+}$ or $\chi_{-}$ fixed points, with the incoming current respectively flowing clockwise or counterclockwise into one of the adjacent wires, depends on the direction of the threaded magnetic field into the ring.~\cite{Oshikawa06} (This point will become more apparent in the next section when discussing Fermi-liquid leads) When $g_1=g_2=g_3=g$, the scaling dimensions listed in Table~\ref{tab:summary-Y-junction} for both $\chi_{\pm}$ fixed points reduce to $\frac{4g}{3+g}$, and hence $\chi_{\pm}$ fixed point is stable for $1<g<3$, which is consistent with what found in Ref.~\onlinecite{Oshikawa06}.  Again, we obtain the stability region of the chiral fixed points through $\Delta_{\mathcal{O}_B}^{\chi_{\pm}} >1$. In Fig.~\ref{fig:3-wires-fixed-point}, such region is shown in orange.

The conductances for chiral-$\chi_{\pm}$ fixed points are, in turn, give by
\begin{equation}
G_{jk}^{\chi_{\pm}}=
2\frac{e^2}{h} \frac{ g_j( g_1+g_{2}+g_{3}) \delta_{jk} + g_j g_k (\mp g_m \epsilon_{jk} -1 )}{g_1g_2g_3+g_1+g_2+g_3} ,
\end{equation}
where $\epsilon_{j,j\pm 1}=\pm 1$ while $\epsilon_{jk}=0$ for $j=k$, and $g_m\neq g_j,g_k$.

\begin{center}
\textbf{d. Asymmetric fixed points}
\end{center}

The asymmetric-$A_i$ fixed points have the property that wire $i$ is decoupled from the junction. Such fixed points are a new feature of a system with unequal Luttinger parameters. As shown in Table ~\ref{tab:summary-Y-junction} and mentioned earlier, the scaling dimensions of the leading irrelevant operators of $A_i$ fixed points are inverse to those of the decoupled, Dirichlet and chiral fixed points. Hence, the stability regions of $A_i$ fixed points are complimentary to those of the other fixed points, and the entire parameter space is covered by at least one of the stable fixed point presented in this work. In Fig.~\ref{fig:3-wires-fixed-point}, the stability regions of $A_{1,2,3}$ fixed point are shown in yellow, gray, and blue, respectively. The regions where two asymmetric fixed points overlap are shown in white.

The conductances of the asymmetric fixed points $A_{i}$ are give by
\begin{equation}
G^{A_i}_{jk}= g_{e}^{i+1,i-1} \frac{e^2}{h} (-1 +\delta_{ij}+ \delta_{ik}+ 2\delta_{jk} -3\delta_{ij}\delta_{ik}),
\end{equation}
where $g_{e}^{m,n}$ is the effective Luttinger parameter for the pair of wires $m$ and $n$. To give an idea of the properties of asymmetric fixed points, we hereby write the conductance tensor explicitly for $A_1$,
\begin{equation}
G^{A_1}= \frac{2 g_2 g_3}{g_2 +g_3} \frac{e^2}{h}
\left(
\begin{array}{ccc}
0 & 0 & 0 \\ 
0 & 1 & -1 \\ 
0 & -1 & 1
\end{array} 
\right).
\end{equation}

The stability of the different fixed points in the $g_{1,2,3}$ parameter space is inferred from the scaling dimensions of Table ~\ref{tab:summary-Y-junction}, and shown in Fig.~\ref{fig:3-wires-fixed-point}. Since each stable fixed point implies a phase of the junction, we shall also refer to the graph defining the regions of stability as a phase diagram. It is convenient to illustrate this three-dimensional phase diagram by some cross sections. As all the scaling dimensions of leading irrelevant operators have a cyclic symmetry on wire indices, the stability regions show a threefold rotation symmetry around the $g_1=g_2=g_3$ axis. Thus, a natural choice for these cross sections is given by planes normal to the equal-$g$ axis: $(g_1+g_2+g_3)/3=\bar{g}$, where the parameter $\bar{g}$, the average of the three Luttinger parameters, labels each cross section. Excluding negative $g_i$, these cross sections are equilateral triangles shown in Fig.~\ref{fig:3-wires-fixed-point}.

We first notice that the decoupled fixed point becomes predominant when $\bar{g} < 2/3$, in which none of other fixed points are stable. One can also show that chiral fixed points appear when $\bar{g}>1$ and the Dirichlet fixed point only appears when $\bar{g}>3$. From Figs.~\ref{fig:3-wires-fixed-point}(b), \ref{fig:3-wires-fixed-point}(d), and \ref{fig:3-wires-fixed-point}(f), we observe that the decoupled, chiral and Dirichlet fixed points are realized around the equal-$g$ axis for $\bar{g}<1$, $1<\bar{g}<3$ and $\bar{g}>3$, respectively. These results are consistent with a junction of three identical TLL wires (indicated as black points at the center of triangles), and show how far the Luttinger parameters can deviate from equal-$g$ axis before these phases break down. We find that, in most regions, these phases are stable under a small perturbation away from the equal-$g$ line. This is relevant for the realization of these phases experimentally, as the TLL wires attached to a junction are likely nonidentical.

As shown in Figs.~\ref{fig:3-wires-fixed-point}(c) and \ref{fig:3-wires-fixed-point}(e), the asymmetric fixed points become important around two marginal points $g_{1,2,3}=1,3$. As a small deviation of Luttinger parameters from these two points easily realizes and switches between asymmetric fixed points, the resultant fixed points are thus highly sensitive to all the three Luttinger parameters, and not just the averaged Luttinger parameter $\bar{g}$. Therefore, precise control over the TLL wires become essential around these points. Note that it may be possible to alter between different asymmetric fixed points, and form a nano-switch if one can tune the Luttinger parameters. We mention in passing that the Luttinger parameters of wires can, in principle, be modified by an external gate capacitively coupled to the wire.~\cite{Safi99b}

Finally, it is worthwhile to compare our findings with those of Aristov and W\" olfle,~\cite{Aristov11b,Aristov12} where two identical wires are connected to a wire with unequal Luttinger parameter, by setting $g_1=g_2=g$. In the repulsive and weak attractive interaction regime, $g \approx g_3 < 3$, in which the N, $\chi_{\pm}$ and $A_i$ fixed points predominate, our results show excellent agreement with their findings. In the strong attractive interaction regime, $g \approx g_3 > 3$, the D fixed point identified in the present work and in Ref.~\onlinecite{Oshikawa06} (by nonperturbative boundary-conformal-field-theory methods), however, was not found by the perturbative renormalization-group approach of Refs.~\onlinecite{Aristov11b,Aristov12}.

\section{Conductance Renormalization for wires contacted to Fermi-liquid leads}
\label{sec:G-renormalization}

The linear conductances of different fixed points were calculated in Secs.~\ref{sec:conductance-2wire} and \ref{sec:3-wire}. Here, we discuss the effect of attaching the wires to Fermi-liquid leads. Remarkably, we find that the conductance of each fixed point, in the presence of Fermi-liquid leads, renormalizes to values that are independent of the Luttinger parameters. This generalizes the following interesting effect for the TLL quantum wire with Luttinger parameter $g$: When attached to leads, the measured conductance is quantized at $e^2/h$, which is different from $g e^2/h$. This discrepancy has been resolved by Maslov and Stone and by Safi and Schulz in Refs.~\onlinecite{Maslov95,Safi95a}. There, they studied an inhomogeneous Luttinger liquid and concluded that the conductance of a TLL wire will only depend on the Luttinger parameter at the contact. As a Fermi-liquid (metal) contact can be thought of as a TLL with Luttinger parameter $g=1$, the measured conductance becomes simply $e^2/h$.

\begin{figure}
\includegraphics[width=6cm]{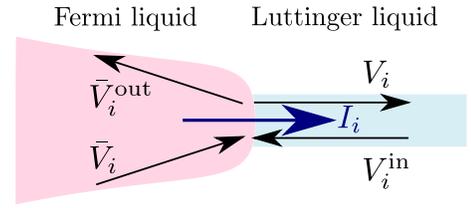}
\caption{The voltages for the incoming and outgoing chiral currents in the Fermi-liquid and connected  TLL wire.}\label{fig:fig6}
\end{figure}
An alternative way to understand this renormalization of conductance due to a Fermi-liquid contact is to introduce a contact resistance, $1/G^{c}= (g-1)(2g) (h/e^2)$, at both contacts in series with the theoretical predicted resistance $1/g (h/e^2)$. This would give the total conductance exactly at $e^2/h$. Here, the contact conductance causes a voltage drop when matching with the wire and gives rise to a current/voltage relation
\begin{equation}
\label{eq:I-V-contact-conductance}
I_i =G^{c}_i (\bar{V}_i-V_i),
\end{equation}
where $V_i$ is the potential of the carriers injected into wire $i$ and $\bar{V}_i$ is the applied voltage at the contact connected to the wire $i$. We can understand this result by thinking in terms of the right- and left-moving currents inside the wire and the Femi-liquid lead as seen in Fig.~\ref{fig:fig6}. The incoming electrons from the Fermi-liquid side are at the voltage $\bar{V}_i$ of the reservoir. The outgoing current may be at a different potential $\bar{V}^{\rm out}_i$ at the contact point, even though the electrons are expected to relax to the equilibrium voltage as they propagate in the lead. Similarly, the right- and left-moving currents in the TLL wire are respectively at voltage $V_i$ and $V_i^{\rm in}$. The junction of the Fermi-liquid lead and the TLL has an effective conductance $g_e e^2/h$ with
\[
1/g_e=\left(1+1/g\right)/2.
\]
The current is then related to the difference in the voltages of the two incoming currents as
\[
I_i=g_e(\bar{V}_i-V_i^{\rm in}).
\]
For the TLL wire, we can similarly write
\[
I_i=g(V_i-V_i^{\rm in}).
\]
Combining the above two equations leads to Eq.~(\ref{eq:I-V-contact-conductance}).
 Generally, it is useful to define a contact conductance tensor $G^c= \delta_{ij} (e^2/h) (2 g_i) /(g_i-1)$ for a junction of three wires.

As the measured conductance is based on the applied voltage at contact, one can define a renormalized conductance tensor, $\bar{G}_{ij}$, as $I_i=\sum_j \bar{G}_{ij}\bar{V}_j$.
To connect this conductance with one we found in the previous subsection, we invoke current conservation:
\begin{equation}
\label{eq:relation-G-Gbar}
I_i=\sum_j G_{ij} V_j= \sum_j \bar{G}_{ij}\bar{V}_j.
\end{equation}
Together with Eq.~\eqref{eq:I-V-contact-conductance}, one can show that~\cite{Oshikawa06}
\begin{equation}
\label{eq:formula-Renorm-conductance}
\bar{G}=(\mathbf{1}+ G G_c^{-1})^{-1} G,
\end{equation}
or equivalently
\begin{equation}
\bar{G}^{-1}=G^{-1}+G_c^{-1},
\end{equation}
which has a simple interpretation of resistances connected in series.

Besides the decoupled fixed point that has obvious vanishing conductance $\bar{G}^{N}=0$, we shall now apply Eq.~\eqref{eq:formula-Renorm-conductance} and obtain the measured conductances of the other fixed points. For Dirichlet fixed point, we have the renormalized conductance
\begin{equation}
\bar{G}_{ij}^D=\frac{2}{3} \frac{e^2}{h} (3\delta_{ij}-1).
\end{equation}
As the largest conductance one can obtain from single-particle unitary scattering is $G^{U}_{jk}= -(4/9)( e^2/h)$ for $j\neq k$,~\cite{Nayak99} the enhanced conductance above demonstrates the role of multi-particle scattering processes.

Upon attaching the wire to external Fermi-liquid leads, the measured conductances of chiral fixed points become
\begin{equation}
\bar{G}_{jk}^{\chi_\pm}= \frac{1}{2}\frac{e^2}{h} \left[(3\delta_{jk}-1) \mp \epsilon_{jk} \right]
\end{equation}
(i.e., the currents flow only from lead $1$ to $2$, $2$ to $3$, and $3$ to $1$ for the $\chi^{+}$ fixed point and in reversed order for $\chi^{-}$ fixed point). Finally, the measured conductance of asymmetric-$A_i$ fixed points reads
\begin{equation}
\bar{G}_{ij}^{A_i}= \frac{e^2}{h} (-1 +\delta_{ij}+ \delta_{ik}+ 2\delta_{jk} -3\delta_{ij}\delta_{ik}),
\end{equation}
which simply indicates a decoupled wire $i$ with the rest of two wires fully conducting.

We note that all renormalized conductance $\bar{G}^{D}$, $\bar{G}^{\chi_\pm}$, and $\bar{G}^{A_i}$, are the same as the unrenormalized conductance $G$ with all $g_i=1$. This result highlights that the dc conductance of a junction of TLL wires depends only on the asymptotic value of the Luttinger parameters of the wires, and in the case when Fermi liquid leads are attached, this asymptotic value is $g_i=1$. Thus, when in contact to Fermi liquid leads, the conductance tensor of the junction will take on the universal values listed above for the different fixed points. Notice, however, that which fixed point is selected still depends on the $g_i$'s of the interacting wire segments.

\section{Conclusion}
\label{sec:conclusion}

In summary, we applied the DEBC method to a two-wire and a Y junction of TLL wires with generally unequal Luttinger parameters. For two spinless wires, we successfully reproduced the prediction that all properties of the junction are determined by a single effective Luttinger parameter $g_e$. We verified this prediction by direct numerical calculations on the lattice with the method of Refs.~\onlinecite{Rahmani10,Rahmani12}: We observed numerically that as long as $g_e<1$, even if one of the wires has attractive interactions $g_i>1$, any impurity leads to a vanishing linear conductance. Moreover, we found that the nonuniversal corrections to the correlations across the junction, which come from perturbations with irrelevant boundary operators to the decoupled fixed point, are independent of the individual Luttinger parameters and only depend on $g_e$ and the local microscopic structure of the junction. For $g_e>1$, we explicitly found a universal conductance of $g_e e^2/h$ regardless of the individual Luttinger parameters and the microscopic details. 

For a Y junction of nonidentical TTLs, we found that the N-BC, $\chi$-BC and the D-BC are stable within regions of the $(g_1,g_2,g_3)$ parameter space, which we explicitly determined. By identifying three more asymmetric fixed points, corresponding to only one decoupled wire, and determining the region of stability of each fixed point, we determined the full phase diagram of the Y junction. We also obtained explicit formulas for the conductance of all fixed points. The findings of this work have direct experimental relevance. In particular, our results shed light on the issue of connecting interacting TLLs to Fermi-liquid leads for the measurement of transport properties. Our work also provides an important theoretical extension of the well-known results on transport through junctions of two and three identical TLL wires.

As an outlook, we finally discuss how to generalize the current method to a junction of $N>3$ wires. The key to such generalization is to identify the possible fixed points via the rotation matrix $\mathcal{R}$. These rotation matrices are constrained by charge conservation and can, in general, be expressed as $SO(N-1)$ matrices after eliminating the total-charge (center-of-mass) mode (cf., Appendix~\ref{app:3-wires}). Then, one has to identify the corresponding N-BC, D-BC, chiral-like BC, and asymmetric BC and analyze their stability. One useful trick for identifying the possible stable fixed points is to utilize the fact that a stable BC would make certain boundary operators effectively equal to identity, c.f. the discussion in Appendix~\ref{app:3-wires}c. Of course, with the number of wires increasing, the number of possible fixed points also increases and the analysis becomes more complicated.

\section*{acknowledgements}
We thank I. Affleck and M. Oshikawa for enlightening discussions, and for collaboration on closely related work. We are also grateful to P. W\" olfle for helpful discussions regarding his results.~\cite{Aristov11b,Aristov12} This work was supported in part by the DOE Grant No. DE-FG02-06ER46316 (A.R., C.C., C.-Y.H.), the U.S. Department of Energy through LANL/LDRD program (A.R.), the NSF Grant No. DMR-0955707 (A.E.F.), and the DARPA-QuEST program (C.-Y.H.).

\appendix

\section{Junction of three wires with unequal Luttinger parameters: DEBC analysis}
\label{app:3-wires}

In this Appendix, we present the details of the DEBC analysis of the stability of the following fixed points for a junction of three quantum wires: decoupled, chiral-$\chi_{\pm}$, Dirichlet, and Asymmetric-$A_{i}$ fixed points. The system is described by the action~\eqref{eq:Bulk-action} with $i=1,2,3$ and the hopping Hamiltonian~\eqref{eq:Hamiltonian-hopping} with $\alpha_{ij}=\gamma/3$, where $\gamma$ is the magnetic flux through the ring at the junction. 

To simplify the notation, we drop the overhead tilde ($\tilde{\;}$) symbol for the rescaled fields throughout this appendix. (All the bosonic fields are rescaled.) To employ the DEBC method, it is convenient to choose a proper basis for the rescaled boson fields. In the first step, we identify the following center-of-mass field, which always satisfies the N-BC due to charge conservation:
\begin{equation}
\Phi_0 = \frac{1}{\sqrt{g_1+g_2+g_3}} (\sqrt{g_1}
\varphi_1+\sqrt{g_2}\varphi_2+\sqrt{g_3}\varphi_3).
\end{equation}
The dual field to $\Phi_0$, i.e.,
\begin{equation}
\Theta_0= \frac{1}{\sqrt{g_1+g_2+g_3}} (\sqrt{g_1}
\theta_1+\sqrt{g_2}\theta_2+\sqrt{g_3}\theta_3) 
\end{equation}
then becomes a constant and can be simply neglected. We then define another two orthonormal boson fields:
\begin{equation}
\begin{split}
\Phi_1 =&\frac{\sqrt{g_2} \varphi_1-\sqrt{g_1} \varphi_2}{\sqrt{g_1+g_2}},
\\
\Phi_2 = &\frac{[\sqrt{g_1 g_3} \varphi_1 + \sqrt{g_2 g_3} \varphi_2 -(g_1 + g_2) \varphi_3]}{\sqrt{g_1+g_2+g_3} \sqrt{g_1+g_2}},
\end{split}
\end{equation}
as well as their dual fields, $\Theta_{1,2}$.
Note that the choice of basis above is arbitrary but, as will become apparent, is a convenient one. We will organize the fields above into a vector $\boldsymbol{\Phi}=(\Phi_1,\Phi_2)^T$ and its dual vector $\boldsymbol{\Theta}=(\Theta_1,\Theta_2)^T$.

It is useful to define the following vectors: 
\begin{equation}
\label{eq:def-K-vectors}
\begin{split}
\boldsymbol{K}_1=&(-\frac{\sqrt{g_1}}{\sqrt{2 g_2
(g_1+g_2)}},\frac{\sqrt{g_1+g_2+g_3}}{\sqrt{2 g_3(g_1+g_2)}}), 
\\
\boldsymbol{K}_2=&(-\frac{\sqrt{g_2}}{\sqrt{2 g_1
(g_1+g_2)}},-\frac{\sqrt{g_1+g_2+g_3}}{\sqrt{2 g_3(g_1+g_2)}}),
\\
\boldsymbol{K}_3=&(\frac{\sqrt{g_1+g_2}}{\sqrt{2 g_1 g_1}},0),
\end{split}
\end{equation}
which will further simplify the notation. Notice that these three $\boldsymbol{K}_i$ vectors add up to $0$ for any $g_i$.

By using the notation in Eq.~\eqref{eq:def-Tjiba} and neglecting $\Theta_0$ (due to the N-BC on $\Phi_0$), the boundary operators for single-particle processes are categorized in four classes, and listed in Table~\ref{tab:Y-BO}. As in the two-wire case, the higher-order processes can be constructed from these single-particle boundary operators.

\begin{center}
\begin{table}
\caption{The boundary operators corresponding to single-particle processes at the Y junction. \label{tab:Y-BO}} 
\begin{tabular*}{\linewidth}{@{\extracolsep{\fill}}*{3}{c}}
\hline\hline 
&
\textbf{(a) $ \pm $ cycle} & \bigstrut
\\ 
\hline
&$T_{21(12)}^{RL} \sim e^{ \pm i \boldsymbol{K}_3
\cdot \boldsymbol{\Phi}} e^{ i
\frac{\sqrt{g_1g_2g_3}}{\sqrt{g_1+g_2+g_3}}(\hat{z}\times
\boldsymbol{K}_3) \cdot \boldsymbol{\Theta}}$& \bigstrut[t] 
\\ 
& $T_{32(23)}^{RL} \sim e^{ \pm i \boldsymbol{K}_1
\cdot \boldsymbol{\Phi}}  e^{ i
\frac{\sqrt{g_1g_2g_3}}{\sqrt{g_1+g_2+g_3}}(\hat{z}\times
\boldsymbol{K}_1) \cdot \boldsymbol{\Theta}}$&
 \bigstrut[t]
\\ 
& $T_{13(31)}^{RL} \sim  e^{ \pm i \boldsymbol{K}_2
\cdot \boldsymbol{\Phi}}  e^{ i
\frac{\sqrt{g_1g_2g_3}}{\sqrt{g_1+g_2+g_3}}(\hat{z}\times
\boldsymbol{K}_2) \cdot \boldsymbol{\Theta}}$ &
\bigstrut[t]
\\ 
\hline
&\textbf{(b) Backscattering} &   \bigstrut
\\ 
\hline
&$T_{11}^{RL} \sim  e^{ -i \frac{2\sqrt{g_1g_2g_3}}{\sqrt{g_1+g_2+g_3}}(\hat{z}\times \boldsymbol{K}_1) \cdot \boldsymbol{\Theta}}$ 
& \bigstrut[t] 
\\
& 
$T_{22}^{RL} \sim  e^{ -i \frac{2\sqrt{g_1g_2g_3}}{\sqrt{g_1+g_2+g_3}}(\hat{z}\times \boldsymbol{K}_2) \cdot \boldsymbol{\Theta}}$ & 
\bigstrut[t]
 \\ 
&$T_{33}^{RL} \sim  e^{ -i \frac{2\sqrt{g_1g_2g_3}}{\sqrt{g_1+g_2+g_3}}(\hat{z}\times \boldsymbol{K}_3) \cdot \boldsymbol{\Theta}}$& \bigstrut[t] 
\\ 
\hline
&\textbf{(c) LL-RR processes}& \bigstrut
\\ 
\hline
&$T_{21}^{LL(RR)} \sim e^{ i \boldsymbol{K}_3
\cdot \boldsymbol{\Phi}}  
e^{\pm i \left(\frac{\sqrt{2g_1 g_2} \Theta_1}{\sqrt{g_1+g_2}} +\frac{\sqrt{g_3}(g_1-g_2)\Theta_2}{\sqrt{2(g_1+g_2)(g_1+g_2+g_3)}}\right)}$&
\bigstrut[t] 
\\ 
&$T_{32}^{LL(RR)} \sim e^{ i \boldsymbol{K}_1 \cdot \boldsymbol{\Phi}}
e^{\mp i \left( \frac{\sqrt{g_1 g_2}\Theta_1}{\sqrt{2(g_1+g_2)}} -\frac{\sqrt{g_3}(g_1+2g_2) \Theta_2}{\sqrt{2(g_1+g_2)(g_1+g_2+g_3)}}\right) }$& \bigstrut[t] 
\\ 
&$T_{13}^{LL(RR)} \sim  e^{ i \boldsymbol{K}_2
\cdot \boldsymbol{\Phi}} 
e^{\mp i\left(\frac{\sqrt{g_1 g_2}\Theta_1}{ \sqrt{2(g_1+g_2)}} +\frac{\sqrt{g_3}(2g_1+g_2) \Theta_2}{\sqrt{2(g_1+g_2)(g_1+g_2+g_3)} } \right)}$ & \bigstrut[t] \\ 
\hline \hline
\end{tabular*}
\end{table}
\end{center}

Using the ansatz~\eqref{eq:BC-R-matrix}, we then write a rotation matrix:
\begin{equation}
\label{eq:RotationMatrix}
\mathcal{R}_\xi=\left(%
\begin{array}{cc}
  \cos{\xi} & \sin{\xi} \\
  -\sin{\xi} & \cos{\xi} \\
\end{array}%
\right),
\end{equation}
where $\xi$ is a rotation angle. The rotation matrix above relates $ \boldsymbol{\phi}_L=(\boldsymbol{\Phi}+\boldsymbol{\Theta} )/2$  to $\boldsymbol{\phi}_R=(\boldsymbol{\Phi}-\boldsymbol{\Theta} )/2$. In terms of $\mathcal{R}_{\xi}$, the scaling dimension of boundary operators is given by Eq.~\eqref{eq:scaling-dimension-Rab}. We now proceed to the stability analysis of the four Y-junction fixed points.

\begin{center}
\textbf{a. Decoupled fixed point}
\end{center}

The decoupled fixed point corresponds to the N-BC for $\boldsymbol{\Phi}$ field, and makes the $\boldsymbol{\Theta}$ field a pure number. Therefore, all backscattering processes are effectively identity. The rotation matrix is simply equal to $\mathcal{R}^{N}_{\xi=0} = \openone$. From Eq.~\eqref{eq:scaling-dimension-Rab}, the scaling dimension of an arbitrary operator with the N-BC becomes $\Delta_{\mathcal{O}_B}^N=|\boldsymbol{a}|^2$. The explicit scaling dimensions for the operators in Table~\ref{tab:Y-BO} are listed below.
\begin{center}
\begin{tabular*}{\linewidth}{l@{\extracolsep{\fill}}c}
\hline\hline
 $\mathcal{O}_B$ & $\Delta^{N}_{\mathcal{O}_B}$ (N-BC) \bigstrut 
\\ 
\hline$T_{21}^{RL}$, $T_{12}^{RL}$, $T_{21}^{LL}$, $T_{21}^{RR}$ & $|\boldsymbol{K}_3|^2=(g_1+g_2)/(2 g_1 g_2)$  \bigstrut[t]  
\\ 
 $T_{32}^{RL}$, $T_{23}^{RL}$, $T_{32}^{LL}$, $T_{32}^{RR}$ &  $|\boldsymbol{K}_1|^2=(g_2+g_3)/(2 g_2 g_3)$  \bigstrut[t]  
\\ 
 $T_{13}^{RL}$, $T_{31}^{RL}$, $T_{13}^{LL}$, $T_{13}^{RR}$ & $|\boldsymbol{K}_2|^2=(g_3+g_1)/(2 g_3 g_1)$
\bigstrut[t]  
\\
 \hline\hline  
\end{tabular*}  
\end{center}
Note that condition $\Delta^{N}_{\mathcal{O}_B}>1$ determines the stability region of the N-BC, shown in red in Fig.~\ref{fig:3-wires-fixed-point}.

\begin{center}
\textbf{b. Dirichlet fixed point}
\end{center}

The Dirichlet fixed point corresponds the D-BC on the $\boldsymbol{\Phi}$ field (i.e., $\boldsymbol{\Phi}$ is effectively a constant at boundary). The rotation matrix of D-BC simply reads $\mathcal{R}_{\xi=\pi}^{D} = - \openone$. From Eq.~\eqref{eq:scaling-dimension-Rab}, the scaling dimension of an arbitrary operator with D-BC becomes $\Delta_{\mathcal{O}_B}^N=|\boldsymbol{b}|^2$. Unlike the N-BC, none of the single-particle processes becomes identity with D-BC. However, some of the two-or-more-particle processes do become identity under D-BC, for instance: $T_{21}^{RL}{T_{12}^{RL}}^{\dag}$, $T_{32}^{RL}{T_{23}^{RL}}^{\dag}$, and $T_{13}^{RL}{T_{31}^{RL}}^{\dag}$. This indicates that the Dirichlet fixed point is associated with the Andreev reflection.

Hereby, we only list the scaling dimensions for the $\pm$ cycle as they are the leading irrelevant operators.
\begin{center}
\begin{tabular*}{\linewidth}{l@{\extracolsep{\fill}}c}
\hline\hline
 $\mathcal{O}_B$ & $\Delta^{D}_{\mathcal{O}_B}$ (D-BC)\bigstrut
\\ 
\hline $T_{21}^{RL}$, $T_{12}^{RL}$ & 
$g_3(g_1+g_2)/2 (g_1+g_2+g_3)$  \bigstrut[t] 
\\ 
$T_{32}^{RL}$, $T_{23}^{RL}$ &  $g_1(g_2+g_3)/2 (g_1+g_2+g_3)$ \bigstrut[t]  
\\ 
$T_{13}^{RL}$, $T_{31}^{RL}$ & 
$g_2 (g_3+g_1)/2 (g_1+g_2+g_3)$\bigstrut[t] 
\\  
\hline\hline
\end{tabular*}  
\end{center}
Now, the Dirichlet fixed point is stable only when all these scaling dimensions $\Delta_{\mathcal{O}_B}^{D} >1$. In Fig.~\ref{fig:3-wires-fixed-point}, the stability region of Dirichlet fixed point (D-BC) is painted in green.

\begin{center}
\textbf{c. Chiral-$\chi_{\pm}$ fixed points}
\end{center}

The chiral-$\chi_\pm$ fixed points are defined as follows: the BC corresponding to the $\chi^+$ ($\chi^-$) fixed point would effectively make all operators in $+$ ($-$) cycle equal to identity for all Luttinger parameters. With this in mind, we could derive the following relationship for the rotation angles of the corresponding rotation matrices:
\begin{equation}
 \tan \xi_{\pm}= \pm \sqrt{\frac{g_1+g_2+g_3}{g_1g_2g_3}}, 
\end{equation}
where $\xi_{\pm}$ are the rotation angles of $\chi_{\pm}$-BC, respectively. The rotation matrices are obtained by plugging in the respective rotation angles into Eq.~\eqref{eq:RotationMatrix}. Here, we do not show them explicitly.

Let us first focus on the $\chi_{+}$ fixed points. The scaling dimensions of the single particle processes (excluding $+$ cycle) are:
\begin{center}
\begin{tabular*}{\linewidth}{l@{\extracolsep{\fill}}c}
\hline \hline
$\mathcal{O}_B$ & $\Delta^{\chi_+}_{\mathcal{O}_B}$ \bigstrut[t]
\\ 
\hline
$T_{11}^{RL}$, $T_{23}^{RL}$, $T_{13}^{LL}$, $T_{21}^{RR}$ & $ \frac{2 g_1 (g_2+g_3)}{ g_1g_2g_3+g_1+g_2+g_3}$  \bigstrut[t] 
\\ 
$T_{22}^{RL}$, $T_{31}^{RL}$, $T_{21}^{LL}$, $T_{32}^{RR}$ &  $\frac{2 g_2 (g_1+g_3)}{ g_1g_2g_3+g_1+g_2+g_3}$  \bigstrut[t] 
\\ 
$T_{33}^{RL}$, $T_{12}^{RL}$, $T_{31}^{LL}$, $T_{13}^{RR}$ & $ \frac{2 g_3 (g_1+g_2)}{  g_1g_2g_3+g_1+g_2+g_3} $\bigstrut[t] 
\\
\hline \hline   
\end{tabular*}  
\end{center}
Notice that these scaling dimensions are cyclic in three indices and hence all operators are important for determining the stability of the $\chi_+$ fixed point. 

As for the $\chi_{-}$ fixed point, one can show that all the leading-order operators have exactly the same scaling dimensions listed in the table above. Thus, both $\chi_\pm$ fixed points share exactly the same stability. In Fig.~\ref{fig:3-wires-fixed-point}, the stability region of chiral-$\chi_{\pm}$ fixed point is shown in orange.

\begin{center}
\textbf{d. Asymmetric fixed points}
\end{center}

Although the use of rotation matrices $\mathcal{R}_{\xi}$ is useful for other fixed points, it is most convenient to identify rotation matrices directly in the rescaled boson field, $\tilde{\phi}_{i}^{L,R}$, basis. Because the decoupled wire effectively has the N-BC for itself and the connected wires should follow mutual D-BC, by using Eq.~\eqref{eq:R-ND-BC}, the rotation matrix of asymmetric fixed point, $A_1$, has the form
\begin{equation}
  \label{eq:R-A1}
\mathcal{R}^{A_1}= 
\left(
\begin{array}{ccc}
1 & 0 & 0 \\ 
0 & \frac{g_2-g_3}{g_2+g_3} & \frac{2 \sqrt{g_2g_3}}{g_2+g_3} \\ 
0 & \frac{2 \sqrt{g_2g_3}}{g_2+g_3} & \frac{g_3-g_2}{g_2+g_3}
\end{array} 
\right),
\end{equation}
while those of $A_{2,3}$ can be constructed by permuting the indices in the corresponding matrix elements. 

By using this rotation matrix \eqref{eq:R-A1}, it is straightforward to show that the following single-particle tunneling processes are equal to identity: $T^{RL}_{32}$, $T^{RL}_{23}$, and $T_{11}^{RL}$. In addition, the scaling dimensions of the leading relevant/irrelevant operators read as follows.
\begin{center}
\begin{tabular*}{\linewidth}{l@{\extracolsep{\fill}}c}
\hline \hline
 $\mathcal{O}_B$ & $\Delta^{A_1}_{\mathcal{O}_B}$ ($A_1$-BC)\bigstrut
\\ 
\hline 
$T_{21}^{RL}$, $T_{12}^{RL}$, $T_{21}^{LL}$, $T_{21}^{RR}$, $T_{13}^{RL}$, $T_{31}^{RL}$, $T_{13}^{LL}$, $T_{13}^{RR}$
& $\frac{g_1+g_2+g_3+g_1g_2g_3}{2g_1(g_2+g_3)} $  \bigstrut[t] 
\\ 
$T_{22}^{RL}$, $T_{33}^{RL}$, $T_{32}^{LL}$, $T_{32}^{RR}$ &  $\frac{2 g_2 g_3}{g_2+g_3}$  \bigstrut[t] 
\\ 
$T_{21}^{RL} {T_{12}^{RL}}^{\dag}$, $T_{13}^{RL} {T_{31}^{RL}}^{\dag}$ & $2(\frac{1}{g_1} +\frac{1}{g_2+g_3} )$\bigstrut[t]  
\\  
\hline \hline
\end{tabular*}  
\end{center}
Here, we notice that some leading order operators are two particle processes. To obtain the scaling behaviors of operators near the $A_{2,3}$ fixed points, one can simply permute the indices of the Luttinger parameters with the corresponding operators. In Fig.~\ref{fig:3-wires-fixed-point}, the stability regions of $A_{1,2,3}$ fixed points are shown in yellow, gray, and blue, respectively.

\bibliography{refer-junction.bib}

\end{document}